\documentclass[amsmath,amssymb,prx,reprint,superscriptaddress,tightenlines,twocolumn,showpacs,floatfix,longbibliography,nofootinbib]{revtex4-1}
\pdfoutput=1

\usepackage{graphicx}
\usepackage{dcolumn}
\usepackage{bm}
\usepackage{color}

\definecolor{nblue}{RGB}{28,130,185}
\definecolor{cgreen}{RGB}{76,153,0}
\definecolor{myorange}{RGB}{245,156,74}

\usepackage{hyperref}
\hypersetup{
  colorlinks=true,
  citecolor=cyan,
  urlcolor=cgreen
}

\def \beq{\begin{eqnarray}}
\def \eeq{\end{eqnarray}}
\def \avOne{\bar{\phi_1}}
\def \avTwo{\bar{\phi_2}}
\newcommand{\bq}{\bm{q}}
\newcommand{\ca}{A}
\newcommand{\iu}{{i\mkern1mu}}
\newcommand{\revise}[1]{\textcolor{black}{#1}}

\begin{document}

\title{Scalar Active Mixtures: The Non-Reciprocal Cahn-Hilliard Model}

\author{Suropriya Saha}
\affiliation{Max Planck Institute for Dynamics and Self-Organization (MPIDS), D-37077 G\"ottingen, Germany}

\author{Jaime Agudo-Canalejo}
\affiliation{Max Planck Institute for Dynamics and Self-Organization (MPIDS), D-37077 G\"ottingen, Germany}

\author{Ramin Golestanian}
\email{ramin.golestanian@ds.mpg.de}
\affiliation{Max Planck Institute for Dynamics and Self-Organization (MPIDS), D-37077 G\"ottingen, Germany}
\affiliation{Rudolf Peierls Centre for Theoretical Physics, University of Oxford, Oxford OX1 3PU, United Kingdom}

\date{\today}

\begin{abstract}
Pair interactions between active particles need not follow Newton's third law. In this work we propose a continuum model of pattern formation due to non-reciprocal interaction between multiple species of scalar active matter. The classical Cahn-Hilliard model is minimally modified by supplementing the equilibrium Ginzburg-Landau dynamics with particle number conserving currents which cannot be derived from a free energy, reflecting the microscopic departure from action-reaction symmetry. The strength of the asymmetry in the interaction determines whether the steady state exhibits a macroscopic phase separation or a traveling density wave displaying global polar order. The latter structure, which is equivalent to an active self-propelled smectic phase, coarsens via annihilation of defects, whereas the former structure undergoes Ostwald ripening. The emergence of traveling density waves, which is a clear signature of broken time-reversal symmetry in this active system, is a generic feature of any multi-component mixture with microscopic non-reciprocal interactions. 
\end{abstract}
\maketitle


\section{Introduction}
The study of active matter \cite{gomp20,MarchettiRMP13} has by now permeated across many scientific fields, length scales, and time scales, ranging from the study of catalytic enzymes \cite{agud18a,JaimeRamin19,jee18} and the cytoskeleton \cite{pros15} inside cells, to the collective motion of cells in tissues \cite{aler20} and suspensions of bacteria \cite{soko12,dunk13}, all the way to the flocking of birds \cite{vics95,tone98}. Throughout the years, particular attention has been given to mechanisms which manifestly break equilibrium physics already at the level of single constituents, as is the case for self-propelled agents such as microswimmers in polar active matter, for the extensile and contractile activity of the constituents in nematic active matter, or for the growth and division in living matter. More recently, subtler manifestations of non-equilibrium activity have taken the spotlight, in particular those related to the interactions between the active agents, which typically include effective non-conservative forces such as those arising from actively-generated hydrodynamic flows, chemical fields, or intelligent cognition and response. Examples of these include the collective behavior of self-phoretic Janus colloids \cite{gole12,saha14,SRGnjp19,star18}, or systems with programmable interaction rules mimicking e.g.~visual perception \cite{lave19}, quorum sensing \cite{baue18,fisc20}, or epidemic spreading \cite{Paoluzzi2020}.

\begin{figure*}
	\includegraphics[width=0.99\linewidth]{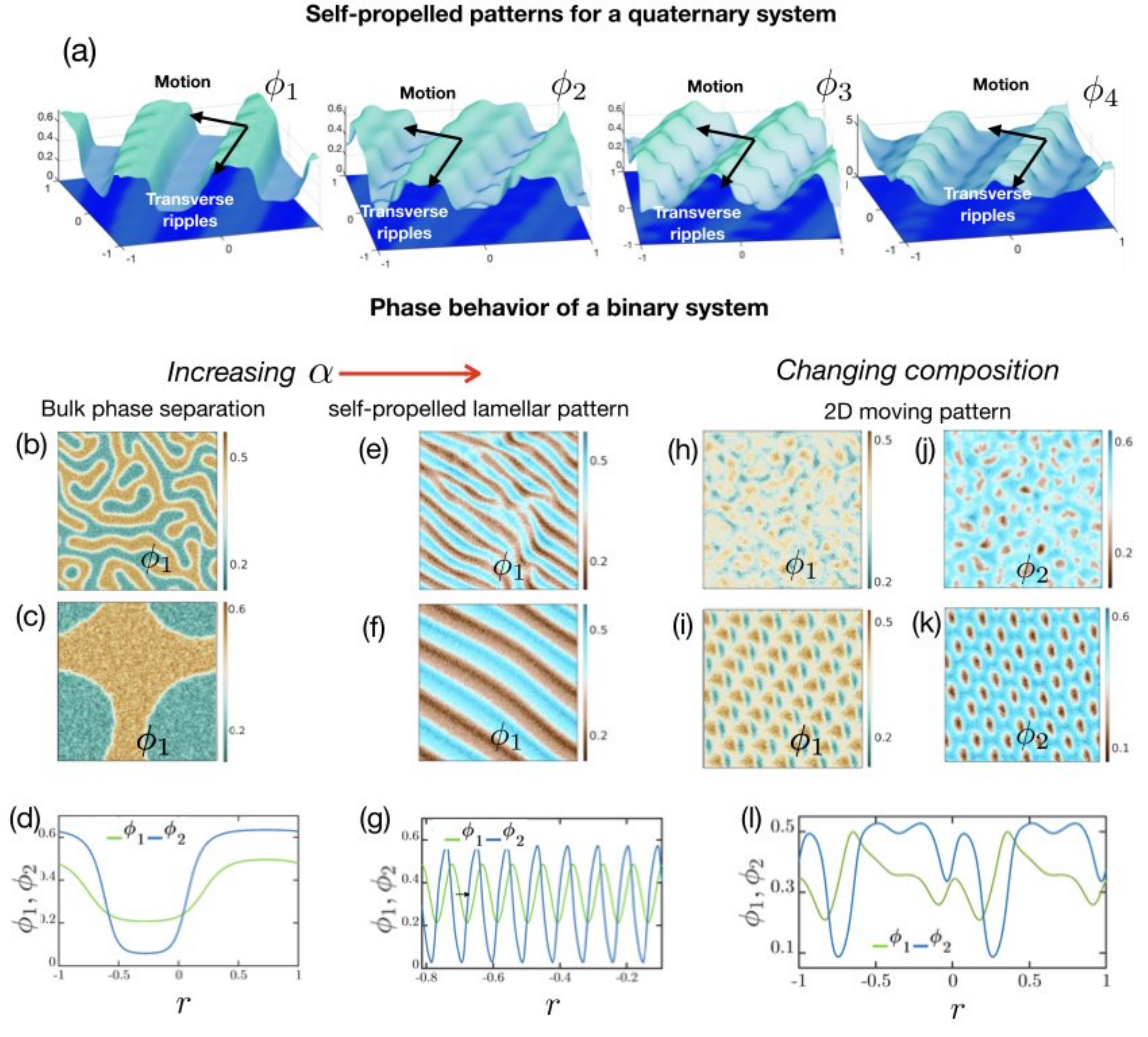}
	\caption{{Phase behavior exhibited by the NRCH model. (a) Example four-component system displaying self-propelled lamellar domains in the steady state. Ripples in density travel perpendicularly to the direction of motion. (b--l) Binary systems can display (b--d) bulk phase separation at low activity ($\alpha=0.2$). At higher activity ($\alpha=0.4$), they undergo an oscillatory instability and display (e--g) a lamellar phase with self-propelled density bands,  or (h--l) two-dimensional moving micropatterns, depending on the system composition. In all cases, the top (b,e,h,j) and bottom (c,f,i,k) rows of snapshots correspond to intermediate and steady states, respectively, in simulations with Gaussian white noise. Line scans of the concentration along a cross section of the corresponding steady states in a noiseless simulation are shown in (d,g,l).  (b) Coarsening of bulk phase separated states proceeds through system-spanning labyrinthine patterns as in equilibrium Cahn-Hilliard, (e) self-propelled lamellar patterns coarsen through annihilation of defects, and (h,j) moving patterns arise from ordering of self-propelled domains. For the self-propelled lamellar and 2D patterns, the concentration profiles at steady state (g,l) show a fixed wavelength and a finite separation between the peaks of the two components, sustaining the motion due to non-reciprocal interactions.  The parameters corresponding to (a) are described in Appendix \ref{app:sim}. For (b--l), we used $c_{1,1} = 0.2$, $c_{1,2} = 0.5$, $c_{2,1} = 0.1$, $c_{2,2} = 0.5$, $\chi = -0.2$, $\chi' = 0$ and average densities $\avOne = 0.35$ and $\avTwo = 0.3$, with the exception of (h--l) for which $\avTwo = 0.42$. \revise{The stiffness $\kappa = 0.0001$ is the same in all simulations. A system size of $201 \times 201$ and a time stepping of $10^{-4}$ are used in all of the simulations presented here.}}}
	\label{fig:Fig1}
\end{figure*}

A particularly interesting realization of active interactions can occur in \emph{mixtures} of non-self-propelling scalar active matter. Here, activity manifests itself only through the nature of the effective interactions between different particle species. Indeed, active particles which are spherically symmetric, for example a fully coated catalytic colloid \cite{SotoRaminPRL14,JaimeRamin19,naso20}, will not self-propel when in isolation and only exhibit anomalous fluctuations \cite{RG2009}. However, when two particles are present, this symmetry is broken and effective interactions between them can arise. For two identical particles, symmetry dictates that the effect that they exert on each other will be reciprocal, with equal magnitude and opposite sign, along the line joining their centers. The active interactions will thus behave like equilibrium interactions, leading to mutual attraction or repulsion.
Crucially, however, if the two particles are different, the effective active interactions are no longer constrained by Newton's third law, and the responses of each particle to the presence of the other need not be reciprocal. This is easily seen in the case of catalytic particles, in which case the phoretic response of a particle of species A to the chemical produced by a particle of species B is generically different from the phoretic response of B to the chemical produced by A \cite{SotoRaminPRL14,JaimeRamin19,naso20,grau20}.
These non-reciprocal interactions can result in substantial departures from equilibrium behavior, such as the formation of self-propelling small molecules \cite{SotoRaminPRL14} or comet-like macroscopic clusters \cite{JaimeRamin19}. However, they can also lead to phase separation into static macroscopic phases with well-defined stoichiometry, not unlike equilibrium phase separation, as was found in our recent work \cite{JaimeRamin19}, in which we explored phase separation in mixtures of chemically active particles interacting through \emph{long-ranged} unscreened chemical fields. Experimentally, such non-reciprocal interactions have recently been observed in a variety of systems composed of isotropic active colloids \cite{niu18,yu18,schm19}.

\revise{Formation of spatial structure has also been reported in multi-component mixtures of {{self-propelled}} active particles with different average speeds  \cite{KolbSoftMatter2020}. Significant departure from equilibrium behavior is observed even with isotropic mutual interactions. Pattern formation, system-wide moving fronts and rafts of active particles are robust observations in such mixtures where oscillatory instabilities are predicted by theoretical calculations. Mixtures of active and passive particles also belong in the same category \cite{WittkowskiNJP2017, StenhammerPRL2015} and show similar behavior. It has been shown using Brownian dynamics simulations that adding even a small fraction of active particles in a passive system leads to departure from equilibrium \cite{Wysocki_2016, StenhammerPRL2015}. Experimental observations in support of these theoretical studies have been reported recently \cite{CuratoloHuandBioArxiv}.}

A number of works in recent years have focused on how non-equilibrium activity enters into continuum theories for scalar active matter, as defined from a top-down approach based on symmetries and conservation laws \cite{hohenbergRMP}, agnostic to the microscopic details of the system. For single-component scalar active matter that satisfies a fluctuation-dissipation relation, it has been shown that activity can give rise to new gradient terms in the dynamical equations that do not come from a free energy, and are responsible for phenomena such as microphase separation and reversed Ostwald ripening, both in frictional \cite{Wittowski14,ClusterBubblyTjhung18} as well as in momentum-conserving systems \cite{tiri15}. On the other hand, keeping the free energy structure of scalar active matter intact but breaking the fluctuation-dissipation relation, while incorporating a density threshold above which diffusivity vanishes, can lead to exotic phenomena such as Bose-Einstein condensation \cite{gole19,maha20}.

For multicomponent scalar active matter, introducing out-of-equilibrium chemical reactions that transform one component into another can also lead to microphase separation \cite{Zwicker2014} and new phenomena such as spontaneous droplet division \cite{RabeaZwicker17}. However, a minimal continuum model that captures the existence of non-reciprocal interactions in scalar active matter has not been proposed yet.

In this work, we take the top-down approach, and introduce a non-reciprocal generalization of the Cahn-Hilliard model. We use continuum equations of motion of the Cahn-Hilliard type and break the equilibrium structure in a minimal way, to study the phenomenon of phase separation in mixtures of scalar active matter interacting through \emph{short-ranged} non-reciprocal interactions. We briefly summarize our results before going into details. In an equilibrium multicomponent system, gradients of thermodynamic chemical potentials drive diffusion of densities to evolve into a bulk separated system of two or more phases. Here, we modify the dynamics of a system of two or more species by adding an extra piece to the chemical potential of species $i$ (or species $j$) by adding a contribution linear in the density of species $j$ (or species $i$) using a coupling constant $\alpha_{ij}$ (or $\alpha_{ji}$), where $\alpha_{ij} = -\alpha_{ji}$. If activity is turned off ($\alpha_{ij} = 0$), each species can still phase separate to coexist in a gas like phase of low density and a fluid like phase of high density. At non-zero $\alpha_{ij}$, non-reciprocal interactions are turned on, and fluid droplets of species $i$ try to co-locate with droplets of species $j$, while the reverse is untrue. In multicomponent systems, this can lead to the formation of very complex moving patterns, such as self-propelling density bands which travel in one direction while smaller density ripples travel in the perpendicular direction, as in the example in Fig.~\ref{fig:Fig1}(a) [see also Movie 1] corresponding to a four-component system.  For mixtures of two components, our model has a single active coefficient $\alpha$. Due to the competition between active interactions and equilibrium reciprocal forces, we find bulk separation [see Fig.~\ref{fig:Fig1}(b--d) and Movie 2] with subtle modifications at low non-reciprocal activity; or oscillatory dynamics leading to either self-propelling bands with an intrinsic wavelength [see Fig.~\ref{fig:Fig1}(e--g) and Movie 3-4] or moving two-dimensional micropatterns [see Fig.~\ref{fig:Fig1}(h--l) and Movie 5-8] at high activity. The observation of system-spanning self-propelling bands is remarkable in that it represents a phase with global polar order arising from a scalar system \emph{via} spontaneous symmetry breaking.  We have explored the full phase diagram of a binary system by varying $\alpha$ and the average composition of the system, using linear stability analysis as well as numerical solution of the equations of motion in 2D. Low activities can modify the bulk phase equilibria and lead to the development of new critical points. Direct transitions between bulk phase separation and oscillatory behavior can be triggered by increasing $\alpha$ beyond a critical value corresponding to an exceptional point; while indirect transitions \emph{via} an intervening homogeneous phase can be obtained by changes in the system composition at constant activity, in which case the system undergoes a Hopf bifurcation. \revise{We note that in parallel to our investigation, You et al. \cite{2005.07684} have examined the effect of non-reciprocity on the dynamics of two coupled diffusing scalar fields, and reported the emergence of traveling bands. Their results complement our work and support the notion that scalar active mixtures with non-reciprocal interactions can generically exhibit time-reversal and polar symmetry breaking, in addition to breaking the time- and space-translation symmetries.}

The paper is organized as follows. In Section~\ref{sec:model}, we present the non-reciprocal Cahn-Hilliard (NRCH) model. We first introduce the multicomponent equilibrium Cahn-Hilliard model, and show how to minimally break its equilibrium structure by adding an antisymmetric matrix of interspecies interaction terms as a non-equilibrium contribution to the chemical potential, followed by a detailed discussion of the binary case. We then begin to describe our results in Section~\ref{sec:bulk}, where we demonstrate the effects of weak non-reciprocal activity on the equilibrium-like bulk phase separation of binary mixtures. In Section~\ref{sec:patternformation}, which contains most of our key findings, we show how at high enough activity, active oscillating phases emerge. This includes a self-propelled active smectic phase consisting of alternating bands of one species which chases after the other species due to short-ranged non-reciprocal interactions, and a phase consisting of a two-dimensional lattice of self-propelled finite-size domains. In Section~\ref{sec:oscillator}, we show how the emergence of oscillations can be linked to the underlying non-conserved dynamics, which are analogous to complex Ginzburg-Landau dynamics with a broken gauge invariance that permits linear oscillations due to the non-reciprocal interactions. The coarsening dynamics of the system towards the active smectic phase, and the associated emergence of global polar order via spontaneous symmetry breaking of the scalar theory, are described in Section~\ref{sec:coarsening}. Finally, we describe the overall phase diagrams for binary mixtures, both in the plane of reciprocal vs non-reciprocal interactions, as well as in composition plane, in Sections~\ref{sec:chialpha} and \ref{sec:patternphases}. We end with a summary and discussion of the implications and future extensions of our work.

\section{Non-reciprocal Cahn-Hilliard model \label{sec:model}}

\subsection{General framework for multicomponent systems}

We consider a system in contact with a momentum sink, due to friction with a substrate for instance, so that the only conservation law is number conservation of each species. Conversion of one species into another is not allowed. The concentrations of the different components are described by fields ${\phi_i}(\bm{r},t)$ with $i=1,\dots,N$ where $N$ is the total number of species. In a passive, equilibrium system, these fields evolve according to the dynamical equations 
\beq
&& \dot{\phi_{i}} + \bm{\nabla} \cdot \bm{j}_i = 0, \nonumber \\
&& \bm{j}_i = - \bm{\nabla} \mu_{i}^\mathrm{eq}   +  \bm{\zeta}_i,
\label{PhiDynamicsEq1}
\eeq
where the current $\bm{j}_i$ of species $i$ includes a contribution from the chemical potential $\mu_{i}^\mathrm{eq}$, as well as from the spatio-temporal Gaussian white noise $\bm{\zeta}_i(\bm{r},t)$ representing fluctuations. The condition of equilibrium implies that the chemical potential can be derived as the functional derivative of a free energy functional $F[\{\phi_i\}]$, as $\mu_{i}^\mathrm{eq} = \delta F / \delta \phi_i$. The free energy must respect the symmetries of the system. The simplest free energy that can be used to describe multicomponent phase separation is the Ginzburg-Landau-type free energy
\beq
F &=& \frac12\int \mathrm{d}\bm{r} \bigg\{ \sum_{i=1}^{N} (\phi_i-c_{i,1})^2(\phi_i-c_{i,2})^2  \nonumber \\ 
&+&2  {\color{black} \sum_{i=1}^{N-1} \sum_{j=i+1}^N \left( \chi_{ij} \phi_i \phi_j + \chi'_{ij} \phi_i^2 \phi_j^2 \right) }  
+  \sum_{i=1}^{N} \kappa_i |\bm{\nabla} \phi_i|^2  \bigg\},\nonumber \\
\label{FreeEnergy}
\eeq
which results in phase separation dynamics of the Cahn-Hilliard type. In absence of any interactions between species, species $i$ phase separates into bulk phases with densities $c_{i,1}$ and $c_{i,2}$ away from the interface. Interspecies interactions stemming from a free energy are controlled by the coefficients $\chi_{ij}$ and $\chi'_{ij}$.

For single-component systems ($N=1$), the different ways in which departure from equilibrium can occur have been studied systematically. For example, Active Model B+ \cite{ClusterBubblyTjhung18} is constructed by supplementing the current $\bm{j}$ in Eq. (\ref{PhiDynamicsEq1}) with additional two leading order terms (in a gradient expansion) in the form of $|\nabla \phi|^2$ and $(\nabla^2 \phi)\nabla \phi$, which cannot be derived from a free energy functional, although they respect the required symmetries. The former term only leads to small modifications to the phase equilibria in the form of macroscopic phase separation, whereas the latter induces strong departures from equilibrium behavior such as microphase separation and reversed Ostwald ripening. Note that, besides the modifications just described which affect the deterministic part of (\ref{PhiDynamicsEq1}), activity can also be introduced by breaking the fluctuation-dissipation relation which links noise and friction in (\ref{PhiDynamicsEq1}).

For active mixtures with $N>1$, however, new ways arise by which the equilibrium structure of (\ref{PhiDynamicsEq1}) can be broken, as we now demonstrate. Most prominently, the presence of several species can lead to non-reciprocal interactions between species. To lowest order in a gradient expansion, we can introduce non-reciprocity by modifying (\ref{PhiDynamicsEq1}) such that
\beq
&& \dot{\phi_{i}} + \bm{\nabla} \cdot \bm{j}_i = 0, \nonumber \\
&& \bm{j}_i = - \bm{\nabla} \mu_{i}^\mathrm{eq} -   \sum_{j} {\alpha}_{ij} \bm{\nabla} \phi_j   +  \bm{\zeta}_i,
\label{PhiDynamicsEq1b}
\eeq
where $\alpha_{ij}$ is a fully antisymmetric matrix, with a total of $N(N-1)/2$ active coefficients for a $N$-component system. The new term proportional to the coupling constants $\alpha_{ij}$ cannot be derived from a free energy and represents the non-reciprocity arising from the activity of the system. We note, however, that the new term can in principle be absorbed into an effective non-equilibrium chemical potential
\beq
\mu_{i}^\mathrm{neq} = \mu_{i}^\mathrm{eq} + \sum_{j=1}^{N} \alpha_{ij} \phi_j,
\label{noneqMu}
\eeq
so that the current of species $i$ can still be written as $\bm{j}_i = - \bm{\nabla} \mu_{i}^\mathrm{neq}   +  \bm{\zeta}_i$. 

The introduction of non-reciprocal activity greatly enhances the space of possible interactions between species \cite{JaimeRamin19}. Indeed, whereas for an equilibrium system with $N$ species the interactions are determined by a symmetric matrix $\chi_{ij}$ with $N(N-1)/2$ independent components (self-interactions are not counted), in the presence of non-reciprocal activity the interactions are determined by the matrix $\chi_{ij}+\alpha_{ij}$ with $N(N-1)$ independent components. To keep the presentation simple, we will therefore focus on binary mixtures with $N=2$ throughout the rest of this paper. As a proof of principle, however, we have simulated a non-reciprocal four-component system; see Fig.~\ref{fig:Fig1}(a) and Movie 1. The multicomponent non-reciprocal interactions lead to a complex oscillatory instability with two competing wavelengths and frequencies, resulting in the formation of self-propelled density bands with smaller density ripples that travel transverse to the direction of propagation of the bands.

\begin{figure*}
	\centering
	\includegraphics[width= 0.99\linewidth]{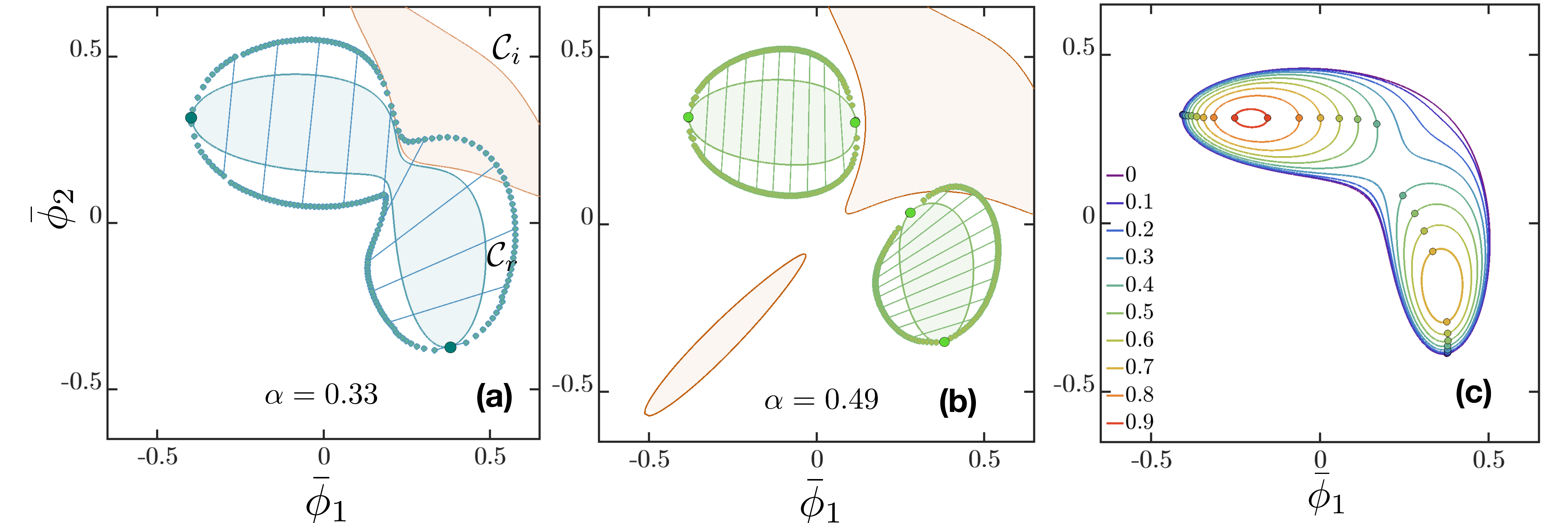}
	\caption{{Bulk phase separation in the plane of average composition $(\avOne, \avTwo)$. (a) The spinodal region, the binodal region, the tie lines and the two critical points are shown for $\alpha = 0.33$. The homogeneous state is unstable to small perturbations inside the spinodal region shaded in blue. (b) At higher activity, $\alpha = 0.49$, the spinodal region has split into two disconnected regions shaded in green. Both regions are surrounded by binodal regions and have developed two new critical points. (c) Change in spinodals and critical points with increasing $\alpha$. Splitting into two disconnected regions occurs for $\alpha>0.35$.  Parameters used in the free energy are $c_{1,1} = 0.2$, $c_{1,2} = 0.5$, $c_{2,1} = 0.1$, $c_{2,2} = 0.5$, $\chi = -0.2$ and $\chi' = 0.2$, $\kappa = 0.001$. \revise{A system size of $201 \times 201$ and a time stepping of $10^{-4}$ are used in all of the simulations used to construct the binodals.}}}
	\label{fig:Fig2}
\end{figure*}

\subsection{Non-reciprocal interactions in a binary mixture}

For a binary mixture, the free energy can be written as
\beq
F &=&  \int \mathrm{d}\bm{r} \bigg\{ \sum_{i=1}^{2} (\phi_i-c_{i,1})^2(\phi_i-c_{i,2})^2  \nonumber \\ 
&& + \chi \phi_1 \phi_2 + \chi' \phi_1^2 \phi_2^2  
+ \revise{\frac{\kappa}{2}}  \sum_{i=1}^{2} |\bm{\nabla} \phi_i|^2  \bigg\},
\label{FreeEnergy2}
\eeq
which results in the equilibrium chemical potentials
\beq
\mu_1^\mathrm{eq} &=& \revise{2} (\phi_1 - c_{1,1})(\phi_1 - c_{1,2})(2\phi_1 - c_{1,1}- c_{1,2}) + \chi \phi_2 \nonumber \\
&& + 2 \chi' \phi_1 \phi_2^2, \nonumber \\
\mu_2^\mathrm{eq} &=& \revise{2} (\phi_1 - c_{2,1})(\phi_1 - c_{2,2})(2\phi_1 - c_{2,1}- c_{2,2})  + \chi \phi_1 \nonumber \\
&& + 2 \chi' \phi_2 \phi_1^2.
\label{ChemPot}
\eeq
We note that, at equilibrium, the sign and strength of the interaction between the two components is governed by $\chi$. If $\chi>0$, the interaction between the two species is repulsive (their overlap increases the free energy of the system), whereas if $\chi<0$, the interaction is attractive (overlap decreases the free energy of the system).

For two components, the activity matrix $\alpha_{ij}$ is simply given by $\alpha_{11}=\alpha_{22}=0$, and $\alpha_{12}=-\alpha_{21}=\alpha$, and there is a single scalar parameter $\alpha$ representing the non-reciprocal activity. The non-equilibrium chemical potentials become 
 \beq
 \mu^\mathrm{neq}_1 = \mu_1^\mathrm{eq} + \alpha \phi_2, \nonumber \\
 \mu^\mathrm{neq}_2 = \mu_2^\mathrm{eq} - \alpha \phi_1.
 \label{ChemPot2}
 \eeq
Considering the form of the equilibrium chemical potentials (\ref{ChemPot}), it becomes clear that the activity $\alpha$ acts to modify the equilibrium interaction parameter $\chi$ within the non-equilibrium chemical potential, so that we find a term $(\chi + \alpha) \phi_2$ in $\mu^\mathrm{neq}_1$, and a term $(\chi - \alpha) \phi_1$ in $\mu^\mathrm{neq}_2$. This implies that the response of one species to the presence of the other becomes non-reciprocal. Suppose that the equilibrium interactions are repulsive, with $\chi>0$, and without loss of generality we take $\alpha>0$. At low activity $0<\alpha<\chi$, the interactions are still repulsive, but species 1 is more strongly repelled from 2 than 2 is from 1. At high activity $\alpha>\chi$, on the other hand, we find that species 1 is repelled from 2, whereas species 2 is attracted to 1.  Similar considerations can be made for a binary mixture which is attractive at equilibrium ($\chi<0$).

\subsection{Linear stability analysis}

In order to obtain more analytical insight into the nature of the instabilities in the system, we linearize the dynamics of the binary NRCH model around a homogeneous state $(\avOne, \avTwo)$ to obtain
\beq 
\begin{pmatrix}
	\dot{\phi_1}(\bq) \\
	\dot{\phi_2}(\bq)
\end{pmatrix}
&=& \begin{pmatrix} \mathcal{D}_{11} & \mathcal{D}_{12} \\ \mathcal{D}_{21} & \mathcal{D}_{22} \end{pmatrix}  \begin{pmatrix}
	{\phi_1}(\bq) \\
	{\phi_2}(\bq)
\end{pmatrix}, \label{eq:matrixeq}
\eeq
where the components of the matrix $\mathcal{D}$ are given by
\beq
\mathcal{D}_{11} &=& -q^2 [ 2 (\avOne - c_{1,1})^2 + 8 (\avOne - c_{1,1}) (\avOne - c_{1,2})  \nonumber \\ && + 2 (\avOne - c_{1,2})^2 + 2 \avTwo^2 \chi'], \nonumber \\
\mathcal{D}_{12} &=& -q^2 [ (\chi+\alpha) +  4 \avOne \avTwo \chi' ],  \nonumber \\
\mathcal{D}_{21} &=& -q^2 [ (\chi-\alpha) +  4 \avOne \avTwo \chi' ],  \nonumber \\
\mathcal{D}_{22} &=& -q^2 [ 2 (\avTwo - c_{2,1})^2 + 8 (\avTwo - c_{2,1}) (\avTwo - c_{2,2})  \nonumber \\ && + 2 (\avTwo - c_{2,2})^2 + 2 \avTwo^2 \chi'], \nonumber
\eeq
In the absence of activity $\alpha=0$, the matrix $\mathcal{D}_{ij}$ is symmetric and thus only admits real eigenvalues. When the non-reciprocal activity is turned on, however, $\mathcal{D}_{ij}$ is no longer symmetric and its eigenvalues may become complex, signaling the possibility of oscillations in the NRCH model.

Indeed, a non-oscillatory instability will take place when one of the eigenvalues $\lambda_{1,2}$ is real and positive, whereas an oscillatory instability is expected when $\lambda_{1,2}$ are a complex conjugate pair with positive real part. To study the phase diagrams of the system, we define $\mathcal{C}_r$ as the region of the parameter space where either $\mbox{Re}(\lambda_1)>0$ or $\mbox{Re}(\lambda_2)>0$, and $\mathcal{C}_i$ as the region where $\mbox{Im}(\lambda) \neq 0$. A non-oscillatory instability will occur in regions of $\mathcal{C}_r$ that do not intersect with $\mathcal{C}_i$, whereas the instability will be oscillatory at the intersection between $\mathcal{C}_r$ and $\mathcal{C}_i$.

\revise{We note that the linearized dynamics in Eq. \eqref{eq:matrixeq} is similar in form to that obtained by an elimination of fast relaxing orientational degrees of freedom for a mixture of self-propelled interacting Brownian spheres \cite{WittkowskiNJP2017}. There is also a strong similarity between Eq. \eqref{eq:matrixeq} and the linearized dynamics reported in \cite{JaimeRamin19}, with the crucial difference that interactions in our model are short-ranged, as can be surmised from the $q^2$ that appears in all terms of $\mathcal{D}$. However, it is also important to note that the oscillatory instabilities arise due to the existence of asymmetry in the dynamical matrix in the cases studied in Refs. \cite{WittkowskiNJP2017}, \cite{JaimeRamin19}, and in the current work, due to different manifestations of nonequilibrium activity. 
These differences become more apparent when the non-linear equations are solved numerically to obtain the steady state.}

\section{Bulk phase separation \label{sec:bulk}}

We now begin to explore the behavior of active binary mixtures in the NRCH model. For the choice of parameters listed in the caption of Fig.~\ref{fig:Fig2}, the curves $\mathcal{C}_r$ and $\mathcal{C}_i$ obtained from linear stability analysis do not overlap for any value of $\alpha$, implying a non-oscillatory instability within the spinodal region enclosed by $\mathcal{C}_r$.  The phase separation behavior is therefore qualitatively similar to that of an equilibrium system, but the activity can have a strong effect on the phase equilibria as well as on the topology and critical points of the phase diagram.
 
 At low $\alpha$, the spinodal $\mathcal{C}_r$ encloses a single connected region, see Fig.~\ref{fig:Fig2}(a).  A system prepared with average composition within the spinodal will coarsen into two macroscopic phases [see Fig.~\ref{fig:Fig1}(b--d) and Movie 2] with compositions dictated by the endpoints of the corresponding tie line, which define the binodal line, as obtained from numerical simulations of the system. The binodal and spinodal lines meet at two critical points.  However, as the activity $\alpha$ is increased, the spinodal and binodal regions shrink until, beyond a critical value, they split into two disconnected regions with the appearance of two extra critical points; see Fig.~\ref{fig:Fig2}(b). Consideration of the topology of the spinodals shows that the splitting into two disconnected regions occurs for $\alpha>0.35$; see Fig.~\ref{fig:Fig2}(c).

\begin{figure}
	\centering
	\includegraphics[width= 1\linewidth]{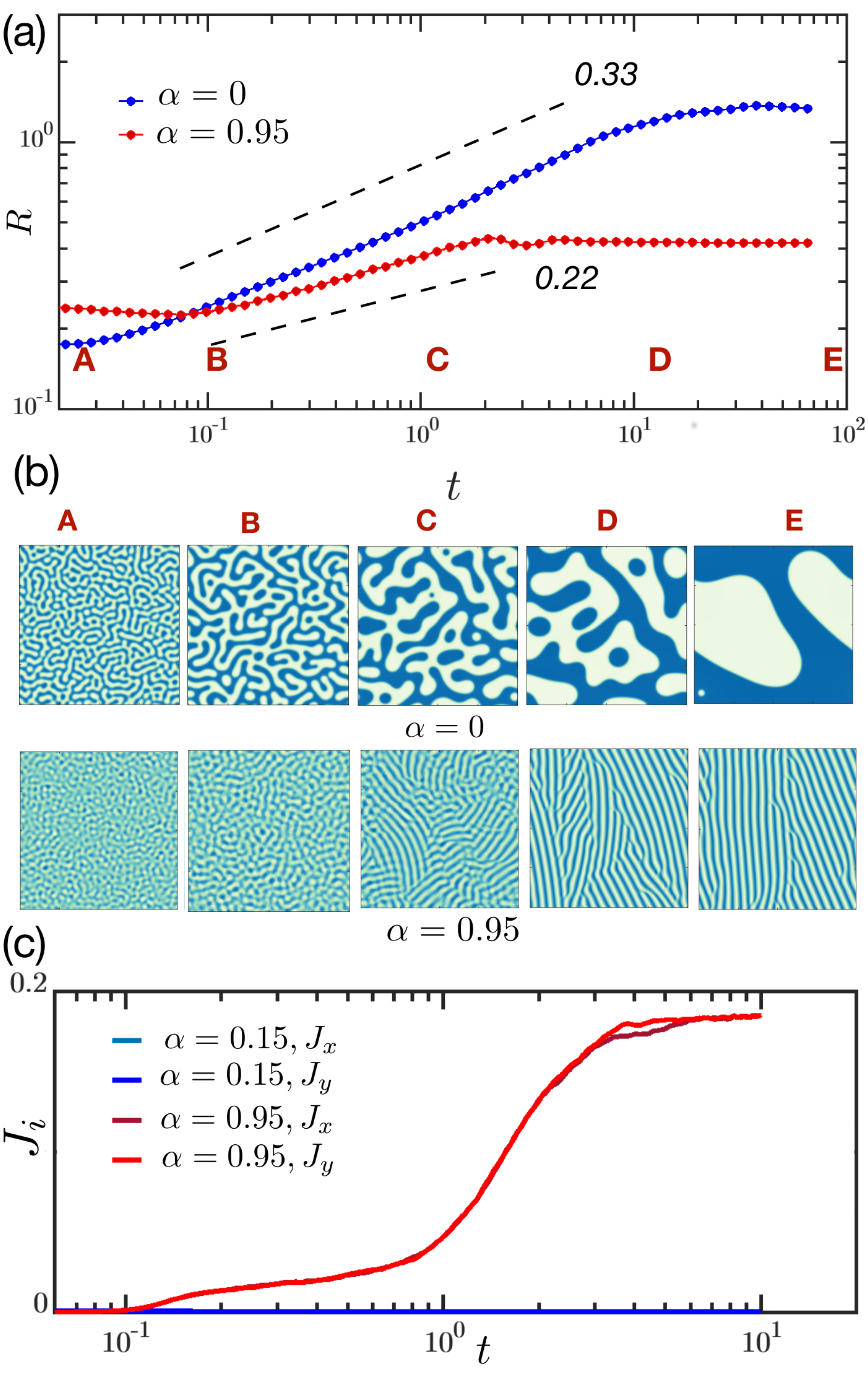}
	\caption{{Dynamics in the NRCH model: (a) Coarsening dynamics for equilibrium bulk phase separation ($\alpha=0$) and formation of self-propelled bands in the active case ($\alpha=0.95$). (b) Panels A-E show simulation snapshots over time. We note the formation of lamellar domains in C which coarsen in D and E to the final state in Fig.~\ref{fig:Fig1}(f). (c) Time evolution of the polar order parameter as measured by the components of the flux $J_x$ and $J_y$, obtained by averaging over $250$ randomly generated \revised{initial} conditions. For $\alpha = 0.95$ where the order parameter saturates to a value $\approx 0.1$ while for $\alpha = 0.15$ it decays to zero. Parameters for the free energy are as in Fig.~\ref{fig:Fig1}. \revise{System sizes of $401 \times 401$ and $201 \times 201$ are used in simulations to calculate the coarsening length scale and current $\bm{J}$ respectively. Time stepping is fixed at $10^{-4}$.}}}
	\label{fig:coarsedyn}
\end{figure}

In an equilibrium system, the binodal lines are determined from balance of chemical potentials and pressure. In our NRCH system we find that, while at steady state the non-equilibrium chemical potentials $\mu^\mathrm{neq}_i$ are balanced in the two phases as expected, the difference in thermodynamic pressure is non-zero indicating that an active contribution balances the equilibrium pressure.
Moving in parameter space along the tie lines by changing the average composition of the system simply extends one phase and contracts the other keeping the composition of each phase unchanged. Lastly, simulations indicate that the coarsening (see Section~\ref{sec:coarsening} and Fig.~\ref{fig:coarsedyn} for details) follows the same growth law as in equilibrium, showing self-similar labyrinthine patterns and an exponent close to $1/3$~\cite{BrayReview94}.

\section{Pattern formation \label{sec:patternformation}}

\subsection{Emergence of self-propelling bands \label{sec:introbands}}

\begin{figure*}
	\includegraphics[width= 1\linewidth]{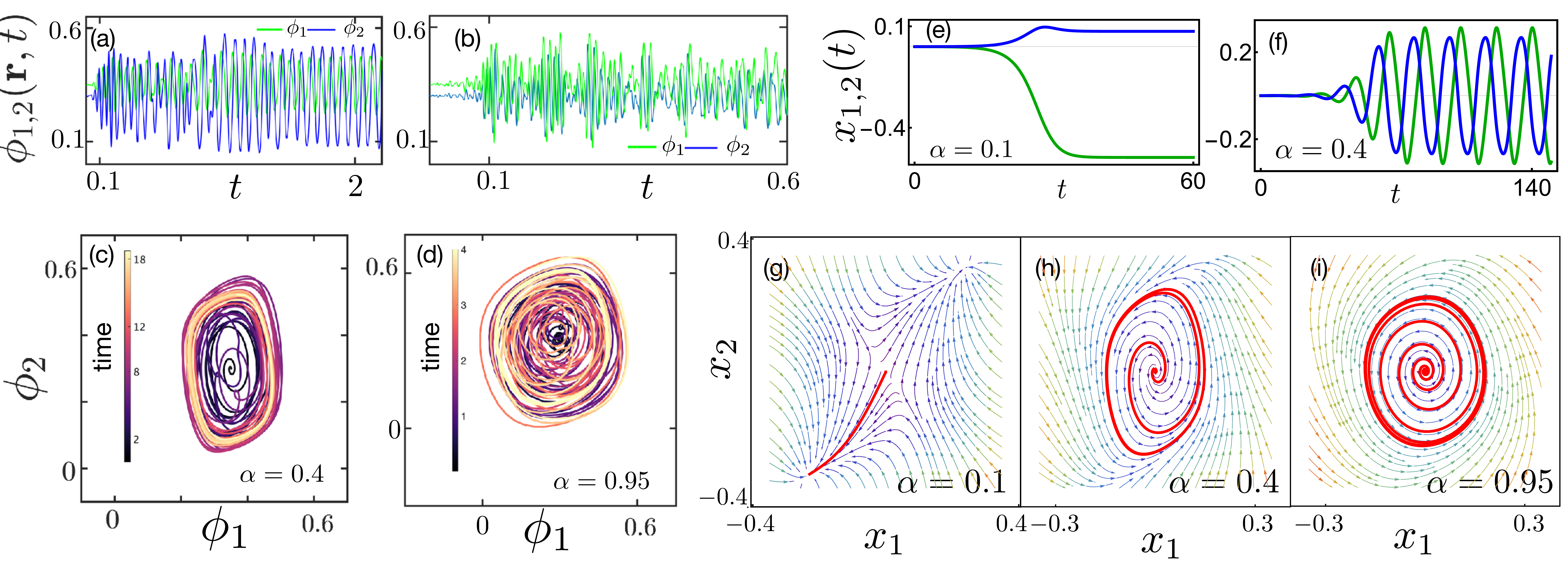} 
	\caption{{Onset of instability and comparison with the \revise{minimal oscillator model described in Sec. \ref{sec:oscillator}}: (a,b) Temporal variation of the density fields $\phi_i$ and (c,d) corresponding phase portraits taken at a randomly chosen point in space $\bm{r}$ for (a,c) $\alpha = 0.4$ and (b,d) $\alpha = 0.95$. Temporal trajectories of the underlying non-conserved dynamical system are shown in (e,f) for $\alpha=0.1$ and $0.4$, with initial conditions very close to $(0,0)$. Phase portraits for $\alpha=0.1$, $0.4$, and $0.95$ are shown in (g--i), with a trajectory originating from $(0,0)$ highlighted in red. (g) At low $\alpha$, there are two stable points. (h,i) Spiral trajectories and a limit cycle arise for large enough $\alpha$. Note the similarity in the form of the limit cycles in panels (c) and (h), and (d) and (i). Parameters for the free energy are as in Fig.~\ref{fig:Fig1}. \revise{The system size of $201 \times 201$ is used in all of the simulations used here, with a time step of $10^{-4}$.} }}
	\label{fig:phaseportrait}
\end{figure*}

For parameters with $|\chi|>\chi'$ and sufficiently large $\alpha$,
we find that $\mathcal{C}_r$ can be contained within $\mathcal{C}_i$, implying an oscillatory instability through which the homogeneous system undergoes active microphase separation. In the steady state, self propelled bands of density of both components are observed to move with constant velocity in a spontaneously chosen direction; see Fig.~\ref{fig:Fig1}(e--g) and Movie 3-4. The maxima of the density profiles of the two species are separated in space, leading to one species chasing the other, as observed for particle dimers \cite{SRGnjp19} or self-propelled clusters \cite{JaimeRamin19} in active colloid systems with non-reciprocal interactions. The densities $\phi_{i}$ oscillate before forming bands that are all oriented in a fixed direction; see Fig.~\ref{fig:phaseportrait}(a,b). This final state, resembling a lamellar phase with several bands of density equivalent to a self-propelled active smectic, is observed when $\alpha$ is increased above a certain threshold, which is $0.23$ for the parameters in Fig.~\ref{fig:Fig1} with composition $(0.35,0.3)$.
As $\avTwo$ is increased at constant $\avOne$, the traveling bands break up into moving micropatterns; see Fig.~\ref{fig:Fig1}(h--l) and Movie 5-8. These micropatterns consist of a lattice of high density domains of each component, again shifted in space leading to effective self-propulsion due to the non-reciprocal interactions.

\subsection{Connection to the underlying oscillator \label{sec:oscillator}}

The route to instability becomes clearer by considering the phase portrait at individual points in space;  obtained by plotting $(\phi_1(\bm{r},t),\phi_2(\bm{r},t))$ for an arbitrary choice of $\bm{r}$; see Fig.~\ref{fig:phaseportrait}(c,d). At short times, the trajectories spiral out from the initial composition converging to limit cycles. Each point in space follows their own path to reach the quasi one-dimensional steady state. 

As discussed above, within the instability line $\mathcal{C}_r$, the mixed state is globally unstable since the dynamical matrix develops a pair of imaginary eigenvalues with positive real parts. Following studies of the complex Ginzburg-Landau equation \cite{Aranson2002}, where the dynamics of the underlying oscillator provides clues to the onset of pattern formation, we study the underlying zero-dimensional system of two variables. As the initial oscillations develop into traveling waves, the spatio-temporal oscillations can be thought of as a field of oscillators in 2D, that are coupled to one another by diffusion gradients. 

To show this, we consider \revise{the {\it{minimal oscillator}}} with {\revise{two degrees of freedom that evolve in time as}} $\dot {x}_i = -\mu^\mathrm{neq}_i (x_1,x_2)$, where the RHS has the same functional form as \eqref{ChemPot2} with the substitution $\phi_i \to x_i$.  At low $\alpha$, the system has two degenerate stable fixed points that are stable nodes with their own basins of attraction; see Fig.~\ref{fig:phaseportrait}(e,g). On increasing $\alpha$, the equations that are linearized around the point $(0,0)$ develop an unstable pair of eigenvalues with non-zero imaginary parts. The corresponding phase portrait resembles a modification of the Hopf bifurcation: trajectories spiral out from the center and converge to periodic limit cycle; see Fig.~\ref{fig:phaseportrait}(f,h,i).  All initial points converge to a limit cycle, oscillating in time with a frequency proportional to $\alpha$. The phase space trajectories of the {\revise{minimal oscillator}} bear strong similarities to those of the corresponding conserved system, as can be seen by comparing Fig.~\ref{fig:phaseportrait}(a,c) to Fig.~\ref{fig:phaseportrait}(f,h), and Fig.~\ref{fig:phaseportrait}(d) to Fig.~\ref{fig:phaseportrait}(i), respectively. \revise{Note, however, that Fig.~\ref{fig:phaseportrait}(e-h) provide a complete representation of the trajectories of a deterministic system with two degrees of freedom whose flowlines cannot cross one another. In contrast, the trajectories in Fig.~\ref{fig:phaseportrait}(a-d) are a two dimensional projection of an infinite dimensional phase portrait. This implies that the latter trajectories can cross each other.}

The comparison to the Ginzburg-Landau dynamics can be taken further if we define a complex field $\ca = \phi_1 + \iu \phi_2$. For a choice of parameters which simplifies the equations of motion,  $\chi' = 0$ and  $c_{1,1} = -c_{1,2} = c_{2,1} = -c_{2,2} = c$, the two-component NRCH can be written in terms of this field as
\beq
\partial_t \ca &=& \nabla^2 \left[ - \left(c^2 + \iu \frac{\alpha}{2} \right) \ca + \frac{3}{8} |A^2|A + \frac{1}{8} A^{*3}  \right]+\kappa \nabla^4 \ca.\nonumber \\
\label{2nNRCH}
\eeq
The most general complex Ginzburg-Landau equation for a complex field $\ca$ is written as \cite{Aranson2002}
\beq
\dot{\ca} &=& - \left[ (1+  \iu a_1 ) \ca - b(1+ \iu a_2) |\ca|^2 \ca \right. \nonumber \\
&& \left. + \kappa (1+ \iu a_3) \nabla^2 \ca  \right] , 
\label{compLG}
\eeq
which can be converted into
\beq
\dot{\ca} &=& - \left[ \ca - b(1+ \iu a_2) |\ca|^2 \ca + \kappa (1+ \iu a_3) \nabla^2 \ca  \right],
\eeq
by using the gauge transformation $\ca \to \exp(-\iu a_1 t ) \ca$. 
(Therefore, non-trivial solutions of the complex Ginzburg-Landau equation stem from a non-zero $a_2$ and $a_3$, while $a_1$ can be set to zero using a gauge transformation.) We observe that this is not true for the NRCH equation \eqref{2nNRCH}, due to presence of the term $\frac{1}{8} A^{*3}$ that breaks gauge invariance. As a consequence, the NRCH model supports oscillations at the linear level which are absent in the complex Ginzburg-Landau equation. We also note that Eq. \eqref{2nNRCH} represents conserved dynamics, unlike the complex Ginzburg-Landau equation.


\subsection{Coarsening dynamics and the emergence of global polar order} \label{sec:coarsening}

Coarsening to the final state proceeds through formation of domains which fuse over time ultimately leading to a steady state in which the bands are aligned, as displayed in  Fig.~\ref{fig:coarsedyn}. The merging of domains occurs through annihilation of oppositely charged unit defects. To study the coarsening dynamics quantitatively, we obtain the domain length $R(t)$ following a standard definition that involves the structure factor
\beq
S(\bq,t)  =\langle \phi_1(\bq,t) \phi_1(-\bq,t) \rangle,
\eeq
and its 2D orientation average $\tilde{S}(q,t) = \int \mbox{d}\theta S(q,\theta,t)$, as follows \cite{Wittowski14}
\beq
R(t) =\frac{ 2 \pi  \int \mbox{d}q \, \tilde{S}(q,t)  }{\int \mbox{d}q \, q \tilde{S}(q,t)}.
\eeq

The variation of this coarsening length scale over time is shown in Fig.~\ref{fig:coarsedyn}. We find a coarsening exponent of 0.22, which deviates from the Lifshitz-Sylozov law that predicts a scaling exponent of $1/3$ \cite{BrayReview94}. Such low coarsening exponents have been previously observed in block-copolymer smectic systems \cite{Harrison2000,Harrison2002}, which display an exponent 1/4, and in stripe patterns as described by the Swift-Hohenberg equation \cite{Elder1992,Cross1995}, with exponents 1/5 and 1/4 in the absence and presence of noise, respectively. The similarity in exponents suggests that, even if the microscopic governing rules for the dynamics of these systems are clearly different (note that the Swift-Hohenberg equation describes non-conserved dynamics), they all share common features in the effective dynamics of the annihilating defects. 

Polar order develops within each lamellar band, as can be quantified using a global order parameter derived from the net flux $\bm{J}$, defined as
\beq
\bm{J} =\left\langle \phi_1 \nabla \phi_2-\phi_2 \nabla \phi_1\right\rangle=\left\langle \frac{1}{2 i}\Big(\ca^* \nabla \ca-\ca \nabla \ca^*\Big)\right\rangle.\nonumber \\
\eeq
The flux is plotted in Fig.~\ref{fig:coarsedyn}. 
In the steady state, $\bm{J}$ is zero for bulk separation and non-zero for self-propelled patterns. The existence of an emergent global polar order in the steady state is remarkable, given the underlying scalar nature of the system. It highlights the strong departure from equilibrium behavior afforded by non-reciprocal interactions.

\begin{figure}
	\includegraphics[width= 0.99\linewidth]{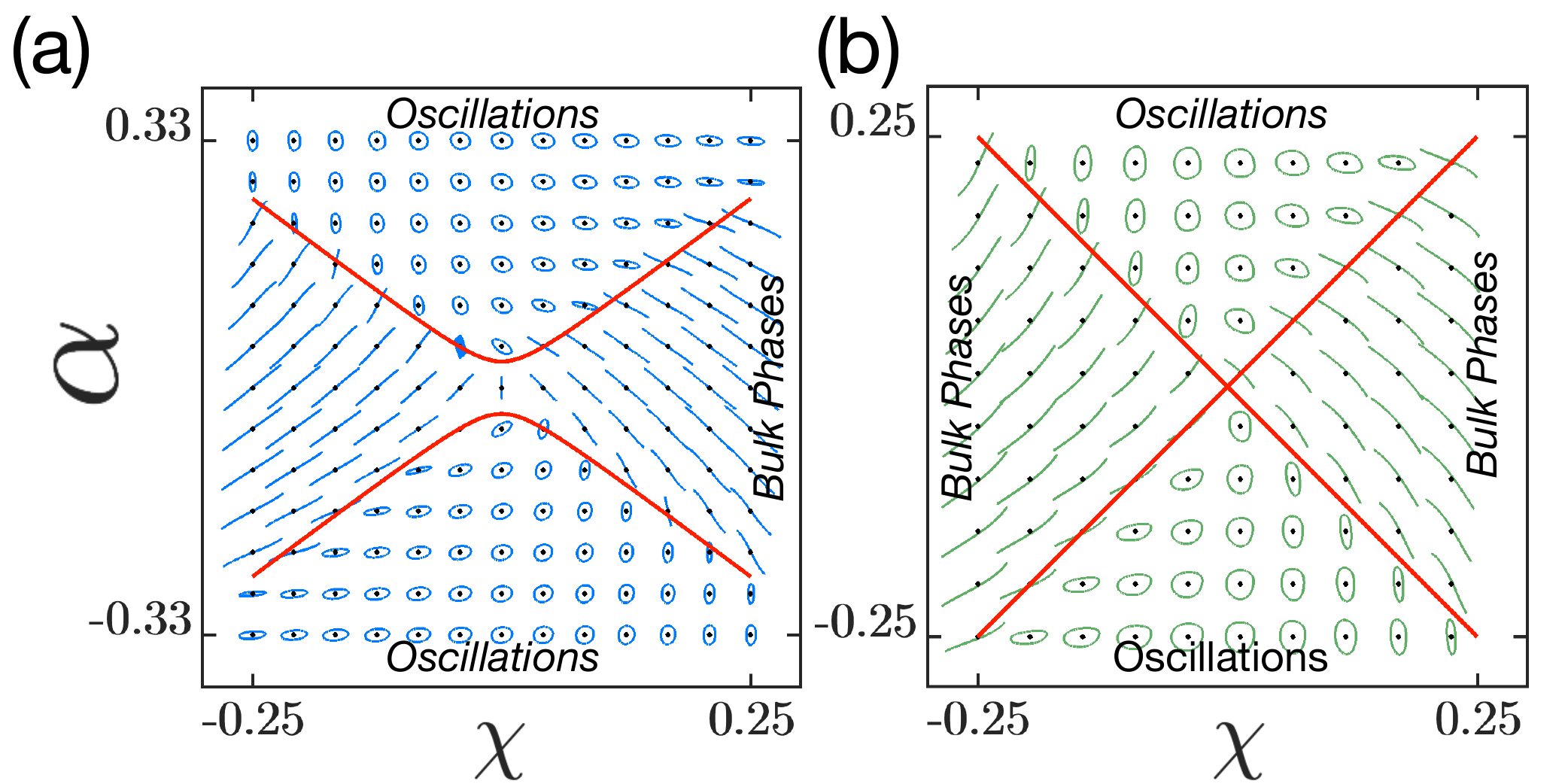}
	\caption{Pattern formation in the plane of reciprocal and non-reciprocal interactions: Stability analysis in the $(\chi,\alpha)$ plane for $c_{1,1} = -c_{1,2} = c_1$ and $c_{2,1} = -c_{2,2} = c_2$, for (a) the general case $c_1 \neq c_2$, and (b) the special case $c_1 = c_2$. The red lines indicate exceptional points at which the eigenvalues collapse and the corresponding eigenvectors become parallel. The results of stability analysis are verified using simulations: the parameters scanned are shown using black dots, and at each dot the steady state limit cycle is plotted to show whether the system goes into oscillations or bulk phase separation. \revise{The stiffness parameter $\kappa = 0.05$. The system size of $201 \times 201$ is used in all of the simulations used here, with a time step of $10^{-2}$.}}
	\label{fig:FigChiDelta}
\end{figure}

\begin{figure*}
\includegraphics[width= 0.99\linewidth]{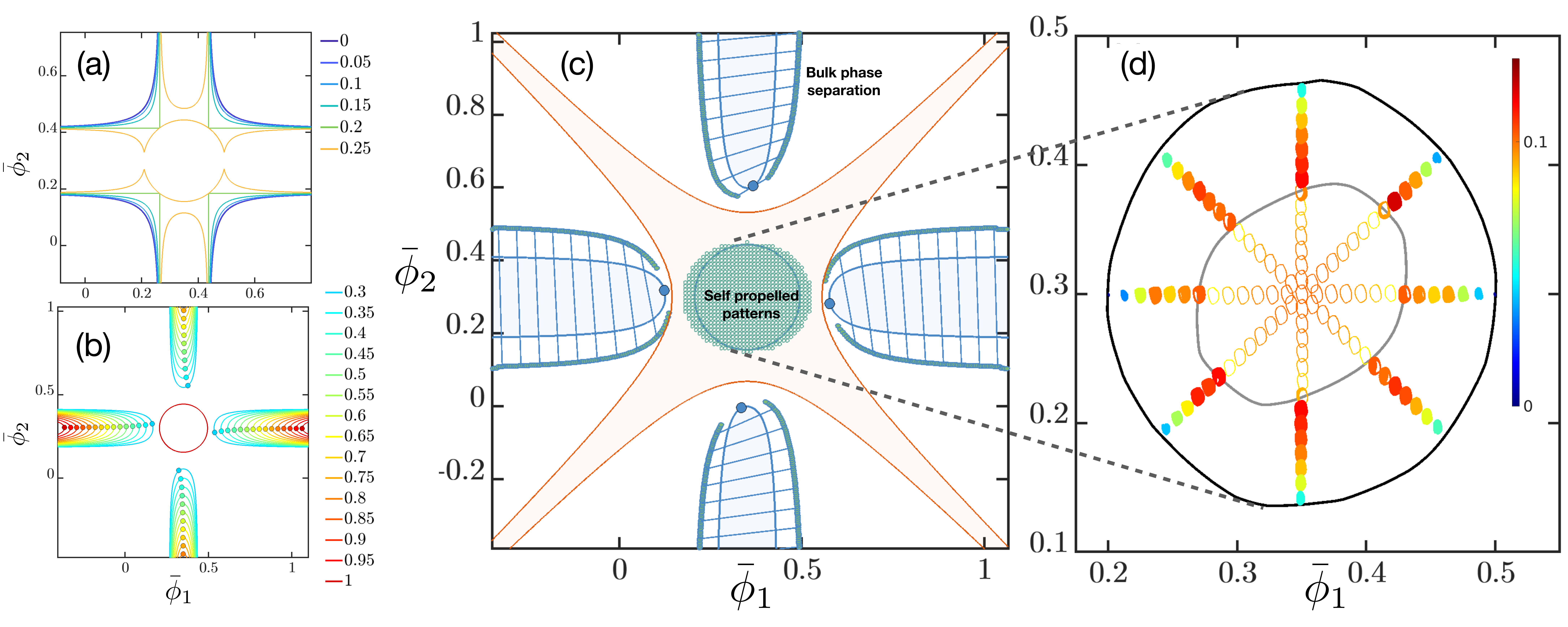}
\caption{{Phase diagrams for pattern formation in the NRCH model. The spinodal region determined from the linear stability analysis is shown in (a) and (b) for the values of $\alpha$ shown in the legend. (a) Spinodals for low $\alpha$. (b) As the activity is increased beyond $\alpha \gtrsim |\chi|$, the topology of the curves consists of four arms surrounding a central circular region. (c) Phase diagram in the plane of average composition $(\avOne, \avTwo)$ for $\alpha=0.4$.  Bulk phase separation (four blue arms) and pattern formation (turquoise inner circle) occur in the same phase diagram. (d) Magnification of the inner circle. The gray line encloses the region where self-propelled lamellar patterns are observed [see Fig.~\ref{fig:Fig1}(e--g)], with a constant wavelength and an amplitude that decays slowly on moving away from the center of the oscillatory region. \revise{Within this region, $(\phi_1(\bm{r},t),\phi_2(\bm{r},t))$ evolve towards a limit cycle independent of $\bm{r}$ at all spatial points, as described in Sec. \ref{sec:oscillator}. The change in the limit cycles upon moving radially along the composition plane are depicted by plotting a scaled down version of Fig.~\ref{fig:phaseportrait}(c) at selected points in a few radial directions. The area of the limit cycle decreases towards the edge of this region; the color encodes the area of the limit cycle and varies as shown in the color bar.} In the region between the gray and black curves, the steady state changes to a lattice of local density undulations which moves with a constant velocity [see Fig.~\ref{fig:Fig1}(h--l)]. \revise{The trajectories change from simple limit cycles to complex repeating patterns. This complexity is reflected in the more complex phase portrait as seen in the scaled down version plotted in the region between the gray and black boundaries.} Parameters for the free energy are as in Fig.~\ref{fig:Fig1}. \revise{The system size of $201 \times 201$ is used in all of the simulations here, with a time step of $10^{-4}$ for panel (c) and an increased time stepping of $10^{-3}$ for the simulations corresponding to panel (d). }}}
\label{fig:Fig4}
\end{figure*}

\subsection{Instabilities in the $(\chi,\alpha)$ plane \label{sec:chialpha}}

We now explore the linear stability of the system for fixed system composition $(\avOne,\avTwo)$ and varying strength of the reciprocal and non-reciprocal interactions, governed by $\chi$ and $\alpha$ respectively. To focus on a particularly simple representative case, we set $\chi' = 0$, such that the equations of motion are now invariant under the shift $\phi_1 \to \phi_1 - (c_{1,1}+c_{1,2})/2$ and $\phi_2 \to \phi_2 - (c_{2,1}+c_{2,2})/2$, and in particular on mixtures with symmetric preferred densities $c_{1,1} = -c_{1,2} = c_1$ and $c_{2,1} = -c_{2,2} = c_2$.  The elements of the dynamical matrix $\mathcal{D}$ linearized around the average composition $(\avOne,\avTwo)=(0,0)$ then become
\beq
\mathcal{D}_{11} &=&  4 q^2 c_1^2, \nonumber \\
\mathcal{D}_{12} &=&  - q^2 (\chi+\alpha), \nonumber \\
\mathcal{D}_{21} &=&  - q^2 (\chi-\alpha), \nonumber \\
\mathcal{D}_{22} &=&  4 q^2 c_2^2.
\eeq
The eigenvalues of this matrix are given by
\beq
\lambda_{1,2} = 2 q^2 (c_1^2 + c_2^2)  \pm q^2 \sqrt{(\alpha_*+\alpha)(\alpha_*-\alpha)}
\eeq
with $\alpha_* \equiv \sqrt{\chi^2 + 4(c_1^2 - c_2^2)^2}$. The corresponding (non-normalized) eigenvectors are
\beq
\eta_{1} = \begin{pmatrix}
\lambda_1 - \mathcal{D}_{22} \\
\mathcal{D}_{21}
\end{pmatrix}
~~\text{and}~~
\eta_{2} = \begin{pmatrix}
\lambda_2 - \mathcal{D}_{22} \\
\mathcal{D}_{21}
\end{pmatrix}. \label{eq:eigenvectors}
\eeq
Because the real part of $\lambda_{1,2}$ is always positive, this implies that the homogeneous state will always be unstable. This instability will become oscillatory when the two eigenvalues collide and become a complex conjugate pair, which gives the condition 
\beq
\alpha^2 \geq \alpha_*^2,
\eeq
for oscillatory behavior. Note that, at the instability, the two eigenvectors become exactly parallel to each other, as can be directly verified from (\ref{eq:eigenvectors}). The minimal value of $\alpha$ beyond which oscillations can occur is $2 |c_1^2-c_2^2|$, which occurs for $\chi = 0$, i.e. when interactions are purely non-reciprocal. The corresponding stability diagram is shown in Fig.~\ref{fig:FigChiDelta}(a), and shows two regions of oscillatory behavior at high positive and negative values of $\alpha$, separated by a gap corresponding to bulk phase separation. This gap vanishes for the singular case $c_1=c_2$, in which case the boundaries between bulk phase separation and oscillations become a pair of lines $\alpha = \pm \chi$, see Fig.~\ref{fig:FigChiDelta}(b). In the oscillatory region for positive $\alpha$, species 2 chases after species 1, whereas the opposite is true for negative $\alpha$. It is also interesting to note that oscillations can occur independently of whether the reciprocal interactions are attractive or repulsive, i.e.~independently of the sign of $\chi$. The bulk phase separated states, on the other hand, have overlapping high-density regions of both components when $\chi < 0$, and non-overlapping high-density regions for $\chi>0$.

The transition lines just described, at which two real positive eigenvalues collide to form a complex conjugate pair with positive real part and the corresponding eigenvectors become parallel, correspond to lines of what are often called \emph{exceptional points} in the non-Hermitian quantum mechanics literature \cite{Kato1995,Heiss_2012}. The coalescence of eigenvalues in this case, which implies parity-time (PT) symmetry breaking, is distinct from degeneracy of eigenlevels in Hermitian quantum mechanics where eigenvectors corresponding to degenerate levels are still non-parallel. Such exceptional points have recently been encountered in other active matter systems such as active solids with odd elasticity \cite{Scheibner2020}.

\subsection{Phase behavior in composition plane \label{sec:patternphases}}

We now consider the phase diagrams in the average composition plane $(\avOne,\avTwo)$ for fixed $(\chi,\alpha)$. For sufficiently high values of the activity with $\alpha \gtrsim |\chi|$, we find that the spinodal splits into five disconnected regions: a middle circular part confined within $\mathcal{C}_i$ and thus corresponding an oscillatory instability, and four arms outside $\mathcal{C}_i$ extending to infinity in four directions, see Fig.~\ref{fig:Fig4}(a,b). The four arms are surrounded by the binodal region where we find bulk phase separation, see Fig.~\ref{fig:Fig4}(c). It is interesting to note that the condition $\alpha > |\chi|$ coincides with the condition required for chasing interactions between the two components to arise, i.e.~for $\chi+\alpha$ and $\chi-\alpha$ to have different sign, as described above.

In the central part of the phase diagram we find rich dynamical behavior. Let us first look at the linear stability analysis. As in the previous section, we focus again for simplicity on the special case with $\chi' = 0$, $c_{1,1} = -c_{1,2} = c_1$ and $c_{2,1} = -c_{2,2} = c_2$. In this case, the equation for $\mathcal{C}_i$ can be written as
\beq
\avOne^2 - \avTwo^2 =  \frac{1}{3} ( c_1^2 +c_2^2 - \chi^2 + \alpha^2 ),
\eeq
which defines a hyperbola,  plotted in orange in Fig.~\ref{fig:Fig4}(c). Inside this curve, where the eigenvalues are a pair of complex conjugates, the curve $\mathcal{C}_r$ is obtained by setting $\mathcal{D}_{11}+\mathcal{D}_{22} = 0$ which yields an equation for a circle with a radius independent of the  value of $\alpha$
\beq
\avOne^2 + \avTwo^2 =  \frac{1}{3}(c_1^2 + c_2^2).
\label{eqCircle}
\eeq 
At this line, which corresponds to the turquoise circle in Fig.~\ref{fig:Fig4}(c), the real part of the eigenvalues crosses from negative to positive values and the system undergoes a Hopf bifurcation, leading to oscillations.

Using simulations, we have investigated in detail the steady-state behavior in this circular region.
The phase space is explored by starting from a single point in the middle and changing the composition along lines emanating radially from this point in uniformly sampled  directions. Our results are summarized in Fig.~\ref{fig:Fig4}(d). The grey line encloses a region where the steady state is the lamellar pattern with a fixed wavelength described in detail above. Between the grey and black lines, the lamellar pattern breaks up into moving two-dimensional micropatterns, see Fig.~\ref{fig:Fig1}(h--l) and Movie 5-8. The amplitude of the limit cycles shrinks as we move outwards towards the edges of the oscillatory region.

\section{Concluding Remarks}

To summarize, we have explored a variety of phases exhibited by scalar active mixtures with non-reciprocal interaction, which form a new class of non-equilibrium phase separation. We find novel oscillations hitherto unreported, in which one component chases after the other due to the non-reciprocal short-ranged interactions. These oscillatory patterns may be effectively one-dimensional, with self-propelled lamellar bands, or they may be fully two-dimensional, resulting in moving lattice-like micropatterns. The lamellar phase constitutes an example of an active self-propelled smectic phase, which remarkably displays global polar order even when the underlying equations of motions have scalar symmetry, and undergoes a very slow coarsening with an exponent around 0.22. The oscillations can be rationalized by considering the non-conserved dynamics of the underlying oscillator, which resemble those of complex Ginzburg-Landau, but with a broken gauge invariance that allows for oscillations at the linear level. Besides these oscillatory regimes, which arise at high values of non-reciprocal activity, we also find bulk phase separation similar to that observed in equilibrium systems, although the composition of the phases is affected by the non-reciprocal activity. We note that active smectic phases have been proposed in the literature for both the apolar \cite{AS2013a} and polar \cite{AS2013b} cases, and a number of specific predictions have been made about their phase behavior, such as \revise{defect-mediated} phase transitions and novel slow-modes. In future work, we aim to perform a systematic study of our emergent active smectic phases to test those predictions. 

Number conservation of each individual component is a defining feature of our model, and one that strongly differentiates it from other multicomponent systems that show oscillatory instabilities, such as reaction-diffusion systems \cite{HalatekFrey18} or Cahn-Hilliard-type models for phase separation coupled to out-of-equilibrium chemical reactions \cite{Zwicker2014,RabeaZwicker17}. In the latter models, there is no number conservation for each individual component (even if the total number may be conserved when components are transformed into each other).

Our NRCH model can be used as a minimal description of multicomponent mixtures of apolar active particles. These could be heterogeneous populations of cells or bacteria communicating through chemoattractants or chemorepellants \cite{kell06,frie09}, solutions of enzymes which participate in common catalytic pathways \cite{zhao18,swee18}, or synthetic systems of catalytically active colloids \cite{niu18,yu18}. In such systems in which particles interact through the concentration fields of a chemical, short-ranged interactions exist when the chemical fields are screened, which is the case in the presence of spontaneous reactions that make the relevant chemical decay, or in the presence of Michaelis-Menten-type kinetics in the production or consumption of the chemicals \cite{saha14}. In the absence of screening, the chemical interactions are long-ranged and can be understood within the framework introduced in Ref.~\citenum{JaimeRamin19}. Besides chemically-mediated interactions, the NRCH can represent apolar particles interacting through thermophoresis, thermally-biased critical mixtures \cite{schm19} or nonequilibrium Casimir forces \cite{Najafi2004}, as well as more complex or intelligent programmed interactions \cite{baue18,lave19}.

Non-reciprocal interactions do not only arise in scalar (apolar) active matter. An example of such interaction in pinned polar active particles with hydrodynamic interactions has been studied in the context of bacterial or ciliary carpets, and shown to be described by an effective frustrated Kuramoto model that cannot be derived from a potential \cite{Uchida2010a,Uchida2010b}. Non-reciprocal interactions also arise naturally in self-phoretic Janus colloids as well \cite{SRGnjp19}, in which case they can show behaviours such as chasing, orbiting, and spiraling. Non-reciprocal interaction should also be commonplace in self-propelled living matter, be it heterogeneous populations of swimming bacteria, or flocks of birds or sheep under attack by a group of predators. Such systems can be approached by considering a multicomponent Vicsek-type model with non-reciprocal alignment interactions between species \cite{Das2002,Maitra2020,fruc20}. We expect that coarse-graining such polar systems into a scalar theory (e.g.~through moment expansion) will result in a scalar theory with non-reciprocal interactions as in our NRCH model, which highlights the usefulness of the top-down approach presented here. Lastly, an open question remains as to how to determine the phase coexistence condition in the NRCH model, in particular regarding the existence of a generalized pressure that is equal in the two phases, as has been obtained for other systems that display active phase separation \cite{GrossbergJoannyPRE15,GenThermoSolon18,ClusterBubblyTjhung18}.

\acknowledgements

We would like to acknowledge stimulating discussions with Philip Bittihn, Beno\^it Mahault, and Yoav Pollack. This research was supported in part by the National Science Foundation under Grant No. NSF PHY-1748958 and NIH Grant No. R25GM067110, as well as the Max-Planck-Gesellschaft. We thank the organizers of the KITP program on active matter for taking the initiative and leadership to run the program virtually during the lockdown period, and for creating a stimulating intellectual environment.


\appendix
\section{Numerical simulations \label{app:sim}}
\revise{The simulations have been performed using a pseudospectral method in two dimensions where the linear terms, which crucially include the fourth order gradient term from surface tension, are treated implicitly in time. By an implicit treatment we mean the following: the pseudospectral method is based on the concept that spectral methods can be used to obtain an exact solution of a linear equation with source terms using a Fourier transformation (which converts the gradients in real space into multiples of wavenumber in Fourier space; e.g. a Laplacian in real space becomes $q^2$ in Fourier space). The nonlinearities in the current contributed by the Cahn-Hilliard free energy, and all the interactions between the components, including the linear cross-diffusion terms, are treated as source terms at every time-step. That is, at every time-step these are evaluated in real-space using values from the previous step to obtain the source terms required to solve for the fields at the current time-step. The pseudospectral method is applicable only for systems with periodic boundary conditions. For the one component Cahn-Hilliard dynamics, implicit treatment of the second and fourth order terms in wavevector, following \cite{Eyre98}, leads to an unconditionally stable algorithm for reaching the ground state with the Euler forward stepping for the time evolution. The multi-component system has to be treated with more care; to this end we run simulations with progressively smaller time steps $\Delta t$ until a convergent steady state is reached. The white noise fields $\bm{\zeta}_{1,2}$ are generated at each time-step from the Gaussian distribution with zero mean and unit width; the forward time stepping is carried out following the standard technique of using $\sqrt{\Delta t}$ as the time increment. The pseudospectral method is implemented in Matlab; the solution of the `minimal oscillator model' and the construction of the spinodals are performed using Mathematica.}

\revise{The system size is chosen to be $201 \times 201$, $401 \times 401$ or $801 \times 801$ in the simulations reported in this work. The system sizes were varied to check for finite size effects. The position-space discretization was carried out using a fixed bin size of $h = 0.01$ for all simulations, so that the length of the system considered is a multiple of $2$, with the coordinates running from $[-1,1]$ for a system size of $201 \times 201$. All simulations reported in this work were carried out with a time step of $10^{-4}$ with the exception of those in Fig. \ref{fig:FigChiDelta} where the step size is $10^{-2}$ and Fig. \ref{fig:Fig4}(d) where the step size is $10^{-3}$.}

\section{Simulation Parameters in Fig. \ref{fig:Fig1}(a) \label{app:para}}
The parameters used for Fig.~\ref{fig:Fig1}(a) are $c_{i,2} = -c_{i,1}$ for all $i$,  $c_{1,1} = 0.2 $, $c_{2,1} = 0.15$, $c_{3,1} = 0.175$, $c_{4,1} = 0.13$, $\chi'_{ij}=0$. The coefficients of the interaction matrix $C_{ij} = \chi_{ij} + \alpha_{ij}$ are chosen as $C_{12} = 0.06$, $C_{13} = -0.04$, $C_{14} = 0.02$, $=C_{21} = -0.07$, $C_{23} = -0.02$, $C_{24} = 0.09$, $=C_{31} = 0$, $C_{32} = 0.11$, $C_{34} = 0.26$, $=C_{41} = -0.01$, $C_{42} = -0.01$ and $C_{43} = 0.05$. The average compositions used are $\bar{\phi}_1 = 0.01$, $\bar{\phi}_2 = 0.02$, $\bar{\phi}_3 = 0.014$ and $\bar{\phi}_4 = -0.07$.

\bibliography{biblio}

\end{document}